\documentclass[%
 reprint,
 amsmath,amssymb,
 aps,
floatfix,
]{revtex4-2}

\usepackage{graphicx}
\usepackage{dcolumn}
\usepackage{bm}
\usepackage{amsmath}
\usepackage{algorithm}
\usepackage{algpseudocode}

\usepackage{bm}
\usepackage{xcolor}
\usepackage{comment}
\usepackage{ulem}
\usepackage{color,soul}

\DeclareMathOperator{\sinc}{sinc}
\newcommand{\dd}{\text{d}}
\newcommand{\deff}{d_{\text{eff}}}

\begin{document}

\preprint{APS/123-QED}

\title{Custom nonlinearity profile for integrated quantum light sources}

\author{Salvador Poveda-Hospital}
\author{Yves-Alain Peter } 
\email{yves-alain.peter@polymtl.ca}
\author{Nicol\'as Quesada }
\email{nicolas.quesada@polymtl.ca}

\affiliation{%
 Department of Engineering Physics, \'Ecole Polytechnique de Montréal, Montréal, QC, H3T 1JK, Canada
}%




\date{\today}

\begin{abstract}
Heralded single-photon sources are a fundamental building block for optical quantum technologies.
These sources need to be unfiltered and integrated to have good scalability and avoid unnecessary losses.
To attain this goal, it is necessary to control the effective nonlinearity seen by the fields as they mix and propagate in a waveguide source.
In this paper, we introduce a method to design nonlinear waveguides with arbitrarily shaped effective nonlinearity profiles.
The method takes advantage of the fact that the second order nonlinear response is a tensor quantity and thus the local effective nonlinearity of a material depends on the propagation direction of the fields participating in the interaction. Thus, by locally changing the propagation direction of the fields we can modulate the wave-mixing process.
Our methods allows for the waveguide fabrication process to be significantly simplified: The material structure of the waveguide is made by a single crystal, no longer needing oriented patterning (OP) or periodic poling (PP).
We use our method to design waveguides with a nonlinearity profile that is Gaussian in the propagation length, allowing to generate perfectly pure heralded single photons. 
\end{abstract}

\maketitle

\section{Introduction}
Optical quantum technologies are poised to grow exponentially in the coming decades~\cite{slussarenko2019photonic}. Computing~\cite{madsen2022quantum}, ultra-precise measurement~\cite{casacio2021quantum}, ultra-fast and secure communication~\cite{yin2020entanglement} are a few of the areas were we expect to see or are already seeing quantum advantage over classical systems. Quantum technologies are still at an early stage, and there are improvements to be made ranging from the generation of photons to the integration of entire systems on a chip.

Spontaneous parametric down-conversion (SPDC), a second order nonlinear process where a photon is fissioned into two lower energy photons, allows us to generate pairs of entangled photons~\cite{kwiat1995new} or pure heralded single photons~\cite{mosley2008heralded}. As quantum technologies evolve, the requirements placed on the generation of quantum states of light become more stringent. The spectral correlation of the photon pairs created must be carefully engineered~\cite{christ2013parametric, grice1997spectral, harder2013optimized}. In addition to introducing spectral requirements, scalability and miniaturization becomes increasingly critical for realizing complex quantum systems~\cite{tanzilli2012genesis, rudolph2017optimistic}. 
Waveguides can be used as photon pairs sources and also to route photons on chips where a huge numbers of components are packaged compactly. They can be engineered to offer good optical confinement, low propagation and coupling losses~\cite{selvaraja2018review}.

If the nonlinear susceptibility is uniform in the waveguide, the phase-matching function (PMF) is given by a $\sinc$ function of the product of the  phase mismatch and half the length of the waveguide~\cite{grice1997spectral}. The side lobes of the $\sinc$ function limit the spectral purity of the biphotons generated in SPDC and motivated U'Ren et al. ~\cite{u2005generation} to study phase-matching functions with better separability properties such as Gaussians. Currently, two methods exist for engineering the joint spectral amplitude of biphotons in waveguides. The first one is spectral filtering, where frequency-selective elements remove the side lobes induced by the $\sinc$ PMF. This method comes at the expense of adding extra losses or lowering efficiencies~\cite{meyer2017limits,blay2017effects}. As noted before, the second method varies the second order nonlinear susceptibility $\chi^{(2)}$~\cite{huang2006amplitude} in step-wise fashion: the sign of $\chi^{(2)}$ can be changed by stacking or growing the nonlinear crystal in different directions~\cite{angell1994growth, eyres2001all}, or by periodically poling the crystal in ferroelectric materials~\cite{miller1964optical, feisst1985current, hum2007quasi}. The modulation of $\chi^{(2)}$ can then be tailored so that the effective nonlinearity has any given functional form along the propagation direction~\cite{branczyk2011engineered, graffitti2017pure, dixon2013spectral, dosseva2016shaping, tambasco2016domain}. This can be used for example so that the effective nonlinearity takes a  Gaussian profile which is the only profile that can give rise to a separable joint spectral amplitude~\cite{quesada2018gaussian}. This modulation leads to an apodized grating that suppresses the side lobes present in $\sinc$ phase-matching function corresponding to flat nonlinearity profile. This technique demonstrated great results on integrated (86\% purity using PP lithium niobate \cite{xin2022spectrally} on a waveguide source) and bulk sources (94\% purity using a PP  potassium titanyl phosphate crystal \cite{kaneda2016heralded}). Nevertheless, if a continuous Gaussian profile could be obtained, the purity could further increase, without reducing the efficiency. Additionally, because no modulation is needed, there is no discretization physical constraint, and then, the signal and idler bandwidth can be much narrower. 
Below, we present a new technique to design waveguides with perfect nonlinear Gaussian profile (or any other profile) and we propose feasible designs using cadmium sulphide (CdS) and gallium phosphide (GaP). Our technique exploits the tensor character of the nonlinear response of a material, in particular the fact that nonlinear wave-mixing processes can be enhanced or suppressed depending on the local propagation direction of the light propagating in the waveguide.
Finally, our technique only needs a thin film layer made of a single crystal structure, contrary to OP where two epitaxial growth with different directions are needed or PP where electrodes need to be deposited. Thus our technique simplifies the fabrication process and eliminates the errors associated to underpoled domains and random variations in domains width which have been shown to damage the spectral purity of heralded single photons~\cite{mann2021low}.
The waveguides have an advantage in terms of reliability, as they do not suffer from the loss of polarization associated with the periodic poling process, making them a reliable and durable option for long-lasting devices.

Our manuscript is structured as follows: In Sec.~\ref{sec:theory} we introduce some basic notation, specify the propagation problem and derive the fundamental equation that guides the design of waveguide paths. In Sec.~\ref{ref:simulations} we present simulation results for different effective nonlinearity profiles and report  materials where our technique can be used to generate spectrally pure single photons. 
The proposed waveguides will be evaluated with the expression of the average number of generated photons developed in Appendix B.
Finally, conclusions and future directions are presented in Sec.~\ref{ref:conclusions}. 

\section{Theory}\label{sec:theory}
\subsection{Nonlinear optics recap}
Nonlinear Optics (NLO) is the study of the electromagnetic field interactions in a nonlinear media, in which the macroscopic polarization $\bm{P}$ (i.e. the dipole moment per unit volume) responds non-linearly to the electric field $\bm{E}$ \cite{new2011introduction}. In this work we will focus on second-order nonlinear interactions for which we can write the constitutive relation
\begin{align} \label{eq:basicP}
    P_{i} = \varepsilon_{0} \left(\chi^{(1)}_{i,j} E_j + \chi^{(2)}_{i,j,k} E_j E_k \right),
\end{align}
where $\chi^{(1)}$ is the linear susceptibility of the material, $\chi^{(2)}$ is the second order susceptibility dictating three wave-mixing processes and we use Latin indices $i,j,k$ for Cartesian components and thus write, e.g., $E_j$ for the $j$-th Cartesian component of the electric field.

To study the dependence of the macroscopic polarization with respect to the waveguide orientation, we will focus on the classical version of SPDC, namely difference frequency generation (DFG) \cite{liscidini2013stimulated,kulkarni2022classical, helt2012does,christ2013theory,quesada2020theory,triginer2020understanding}. Since the effective nonlinearity profile in the classical process of DFG is exactly the same appearing in SPDC we only need to study the former to make predictions about the later. 

In integrated photonics, waveguides are predominantly rectangular channels. In rectangular channels, a wave guided in the core is either TE or TM, consequently the electric field direction is either horizontal ($e$) or vertical ($o$).

\begin{figure}[h!] 
\centering
    \includegraphics[width=0.45\textwidth]{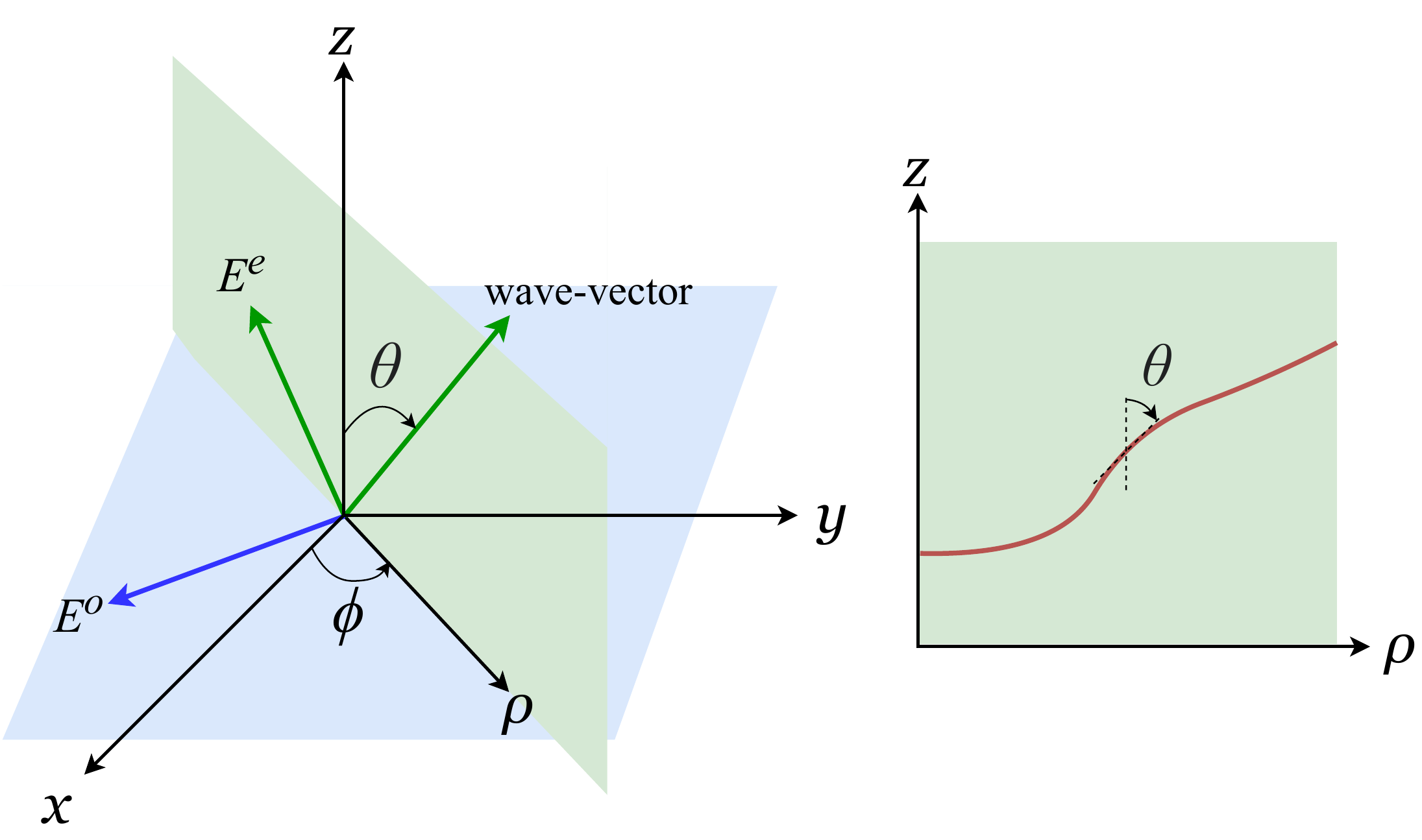}
    \caption{Propagation geometry. We assume the path of the waveguide lies in the green plane. This plane is uniquely specified by the fact that it contains the $z$ axis and makes an angle $\phi$ with the $x$ axis.}
    \label{fig:ThetaPhi_axis}
\end{figure}

The waveguide/wave-vector direction is defined by the angles $\phi$ and $\theta$, depicted in Fig.~\ref{fig:ThetaPhi_axis}.
In practice, the angle $\phi$ is fixed by the crystal orientation. 
The angle $\theta$ defines the waveguide/wave-vector direction in the wafer plane (green plane in Fig.~\ref{fig:ThetaPhi_axis}). 
The value of $\theta$ represents the rotation angle around the normal axis of the wafer plane. The normal of the wafer plane corresponds to $E^o$. 

Relative to the propagation direction, the vertical and horizontal electric fields can then be written as
\begin{subequations}
\begin{align}
	\bm{E}^o &= \label{eq:Eo}
	\begin{pmatrix}
		\sin (\phi) \\
		-\cos (\phi)\\
		0 \\
	\end{pmatrix}
	|\bm{E}^o|, \\
	\bm{E}^e &=
	\begin{pmatrix}
		-\cos (\theta)~ \cos (\phi) \\
		-\cos (\theta)~ \sin (\phi)\\
		\sin (\theta) \\
	\end{pmatrix}
	|\bm{E}^e| ~,
\end{align}	
\end{subequations}
where $|\bm{E}| = \sqrt{E_1^2+E_2^2+E_3^2}$.
There are three possible wave mixing interactions of the electric field guided in the waveguide:
\begin{subequations}\label{eq:polarization}
\begin{align}
\label{eq:E1ee} 
\bm{P}^{ee} (\phi, \theta) &= 2\varepsilon_0 ~
\bm{d}
\begin{pmatrix}
\cos^2 (\theta)~ \cos^2 (\phi) \\
\cos^2 (\theta)~ \sin^2 (\phi)\\
\sin^2 (\theta) \\
-\sin (2\theta)~ \sin (\phi) \\
-\sin (2\theta)~ \cos (\phi) \\
~~\cos^2 (\theta)~ \sin (2\phi) \\
\end{pmatrix}
|\bm{E}^e||\bm{E}^e| ~,\\
\label{eq:E1oo} 
\bm{P}^{oo} (\phi, \theta) &= 2\varepsilon_0 ~
\bm{d}
\begin{pmatrix}
	\sin^2 (\phi) \\
	\cos^2 (\phi)\\
	0 \\
	0 \\
	0 \\
	-\sin (2\phi) \\
\end{pmatrix}
|\bm{E}^o||\bm{E}^o| ~,\\
\label{eq:E1eo} 
\bm{P}^{eo} (\phi, \theta) &= \bm{P}^{oe} (\phi, \theta) = \\
& \quad \quad \quad 2\varepsilon_0 ~ \bm{d} 
\begin{pmatrix}
	- \frac{1}{2} \cos (\theta) \sin (2\phi) \\
	~ \frac{1}{2} \cos (\theta) \sin (2\phi) \\
	0 \\
	- \sin (\theta) \cos (\phi)  \\
	\sin (\theta) \sin (\phi)  \\
	\cos (\theta) \cos (2\phi)  \\
\end{pmatrix}
|\bm{E}^e||\bm{E}^o| ~. \nonumber
	\end{align}
\end{subequations}
In the last set of equations we took advantage of the Kleinman permutation symmetry of the $\chi^{(2)}$ tensor to use contracted notation where we write the $3 \times 6$ matrix $d_{i,jk} = \tfrac12 \chi^{(2)}_{i,j,k} $ (cf. Sec. 1.5.6 of Boyd~\cite{boyd2020nonlinear}). 

Note that the interaction analysis in Eq.~\eqref{eq:polarization} was first done to find phase matching conditions in bulk crystals by Midwinter and Warner in 1965 \cite{midwinter1965effects}. However, their results reflect the projection of the macroscopic polarization to the horizontal and vertical directions, therefore, their results cannot be extrapolated to integrated photonics since not all the wave-vector directions are guided by the waveguide. 

Having described the electric field interactions in a waveguide, it is possible to calculate the non-linear macroscopic polarization for different crystals. 
The interaction in Eq. \eqref{eq:E1oo} is not useful for using the in-plane orientation of the waveguide for controlling the effective nonlinearity since it only depends on the azimuthal angle $\phi$, which only specifies the orientation of the plane where the waveguide is printed.
Notice that, unlike apodized gratings, where $\chi^{(2)}$ is simplified to a 1D vector, in our analysis we use the full tensor nature of $\chi^{(2)}$ (or rather its contracted version $d$).

The analysis just presented allows us to identify situations where it is possible to modify the effective nonlinearity via the orientation of the waveguide. Consider for example Eq.~\eqref{eq:E1eo} for gallium arsenide for which we have~\cite{boyd2020nonlinear}
\begin{align}
\bm{d} = \begin{pmatrix}
    0 & 0 & 0 &d_{14} &0 &0 \\
    0 & 0 & 0 &0 &d_{14} &0 \\
    0 & 0 & 0 &0 &0 &d_{14} 
\end{pmatrix},
\end{align}
and furthermore assume that the waveguide is in the plane $\phi = 0$ to obtain
\begin{align}
\bm{P} = - d_{14} \sin(2\theta) |\bm{E}^e||\bm{E}^o| \begin{pmatrix} 
	0 \\
	1 \\
	0\\
	\end{pmatrix}.
\end{align}
Note that the macroscopic polarization field generated by the nonlinear interaction corresponds to the vertical ($o$) polarization (cf. Eq.~\eqref{eq:Eo} with $\phi=0$). Thus this particular configuration will generate waves that are guided in the $xz$ plane (corresponding to $\phi=0$).

Table~\ref{tab:summaryAPM} summarizes the resulting electric field for different crystals at different orientations. Only the results where $\bm{P}$ is perpendicular to $\bm{E}^o$ or $\bm{E}^e$ are kept because the end goal is to create entangled photons in two orthogonal polarizations. Borrowing terminology from second harmonic generation, we use ``Type I'' to refer to the $ee$ interaction (Eq. \eqref{eq:E1ee}) and ``Type II'' to refer to the $eo$ interaction (Eq. \eqref{eq:E1eo}) \cite{chekhova2021polarization}. 



\begin{table*}
\def\arraystretch{1.5}
\begin{tabular}{|l|c|c|l|l|}
\hline
\multicolumn{1}{|c|}{\begin{tabular}[c]{@{}l@{}} \textbf{Crystal} \\ \textbf{Class} \end{tabular}} 
    & \textbf{Crystal examples}                                 
    & \textbf{Type}       
    & \multicolumn{1}{c|}{$\phi$\textbf{(rad)}} 
    & \multicolumn{1}{c|}{\textbf{Effective nonlinearity}} \\ 
\hline {mm2}  & LiSe, KTP & {II} & 0   & $-d_{24}~\sin (\theta)$  \\ 
\cline{4-5}  \footnotesize{if $d_{31}=d_{32}$} &  &  & $\pm \pi/2$ & $\pm d_{15}~\sin (\theta)$   \\ 
\hline 3m  & LN, LT, BBO  &  II  &  $\pm \pi/2$  & $\pm d_{31} \sin (\theta) + d_{22} \cos (\theta)$\\ 
\hline 32  & Te, $\text{SiO}_2$   &  I  & $\pm \pi/2$  & $-d_{11} \cos ^2 (\theta) \pm d_{14} \sin (2\theta)$ \\
\hline m (m$\perp x$) &  KNB, RNB  &  II  & $\pm \pi/2$  & $\pm d_{15} \sin (\theta) - d_{16} \cos (\theta)$ \\
\hline m (m$\perp y$) & $\text{BaGa}_4\text{Se}_7$, DAST  &  II  & $0$  & $-d_{24}\sin (\theta) + d_{12} \cos (\theta) $ \\
\hline {$\bar{4}2$m} & ADP, KDP, CDA,  & {I}  &  0  & $-d_{14}~\sin (2\theta)$ \\ 
\cline{4-5} & AGSe, AGS &  &  $\pm \pi/2$       & $\mp d_{14}~\sin (2\theta)$  \\ 
\cline{4-5} & CGA, CSP, ZGP &  &  *                 & $-d_{36}~\sin (2\theta)$ \\
\hline {$\bar{4}3$m} & GaAs, GaP, & I & 0  & $ -d_{14}~ \sin (2\theta)  $ \\ 
\cline{4-5} &  InP, InSe, ZnTe  &     & $\pm \pi/2$  & $\mp d_{14}~\sin (2\theta)  $  \\ 
\cline{4-5}23 & $\text{NaClO}_3, \text{NaBrO}_3$        &     &  *           & $-d_{14}~\sin (2\theta)$  \\ 
\hline $\bar{4}$  & $\text{InPS}_4$, $\text{HgGa}_2\text{S}_4$ &  I  & * & $d_{31} \cos(2\theta) - d_{36}\sin (2\theta)$ \\
\hline$\bar{6}2$m & GaSe & I  & 0           & $-d_{22} ~\cos^2 (\theta) $  \\ 
\cline{3-5}       &      & II & $\pm \pi/2$ & $-d_{22} ~\cos (\theta) $  \\ 
\hline {6mm}  & CdS, CdSe,   & {II} & 0   & $-d_{15}~\sin (\theta)$   \\ 
\cline{4-5} &  ZnO    &    & $\pm \pi/2$   & $\pm d_{15}~\sin (\theta)$  \\ 
\hline
\multicolumn{5}{l}{* \footnotesize{Fixing $E^o$ to the $z$ axis of Fig. \ref{fig:ThetaPhi_axis}, the wafer plane is then $(xy)$.}  }
\end{tabular}
\caption{Effective nonlinearity for various crystal classes. Note that effective nonlinearity written in the last column can always be expressed in the canonical form in Eq.~\eqref{eq:F(s)} by using sum ($\sin(x+y) = \sin x \cos y + \cos y\sin x$) and double angle ($\cos 2 x = \tfrac12[1+\cos^2 x]$) trigonometric identities. }
\label{tab:summaryAPM}
\end{table*}

\subsection{Custom effective susceptibility profile}

As shown in Table~\ref{tab:summaryAPM}, for many materials, it is possible to tailor the effective nonlinearity and consequently, as we will show, the waveguide orientation angle $\theta$ can be varied in order to obtain any effective nonlinearity profile. This is because the geometry of the waveguide determines locally the wavevector of the modes. 

Consider for the sake of the argument a material for which we can write 
\begin{align}
\deff = d_0+ d \sin(k \theta+\varphi).
\end{align} 
Note that for the materials in Table~\ref{tab:summaryAPM} we have $k \in \left\{1,2\right\}$ and $|d_0| \leq d$. The latter condition implies that we can always make $\deff = 0$ by setting $k \theta = - \varphi$. Furthermore, whenever $d_0=0$ we can make the effective nonlinearity to take its maximum (minimum) value by making sure that the tangent of the curve  describing the waveguide makes an angle of $\theta = \tfrac{1}{k}\left[\tfrac{\pi}{2} -\varphi \right]$ ($\theta = \tfrac{1}{k}\left[-\tfrac{\pi}{2} -\varphi \right]$) with the $z$ axis. 

More generally, assume that we want to modulate the relative effective nonlinearity so that at position $s$ along the length of the waveguide it takes the value $F(s)$, thus we want the following to hold
\begin{align}\label{eq:F(s)}
\frac{\deff}{d} = \frac{d_0}{d}+ \sin(k \theta+\varphi) = F(s)
\end{align}
where recall, $\theta$ is the angle the tangent to the curve describing the waveguide makes at point $s$, while $F(s)$ is some prescribed target nonlinearity profile. 

Note that we can always find an orientation $\theta$ as long as $-1+\frac{d_0}{d} \leq F(s) \leq 1+\frac{d_0}{d}$. In particular, we can always find an orientation for $|F(s)|\leq 1$ as long as $d_0=0$. Moreover, for the materials listed in Table~\ref{tab:summaryAPM} for which $|d_0|<d$  Eq.~\eqref{eq:F(s)} always has a solution for $\theta$ as long as $0 \leq \text{sign}(d_0) F(s) \leq 1$.

We now formalize the observation presented in the previous paragraph and derive an equation that the path of the waveguide needs to satisfy so that along the propagation direction $s$ the effective nonlinearity takes the value $F(s)$.
Any point in the green plane in Fig.~\ref{fig:ThetaPhi_axis} can be specified by its height along the $z$ axis and its distance to the origin in the $xy$ plane, namely $\rho = \sqrt{x^2+y^2}$.
The path of the waveguide will be given by a curve
\begin{align}
\rho = g(t) \text{ and } z = h(t)
\end{align} 
specified by the parameter $t$. Often it is possible to eliminate the parameter $t$ and write explicitly $z = f(\rho)$. Finally, we use $s$ to specify the position of a point along the curve describing the waveguide with respect to some arbitrary starting point $z_0 = z(t_0), \rho_0 = \rho(t_0)$. The value of $s$ can be found using the standard definition of arc-length
\begin{align}\label{eq:arclenparam}
s = \int_{t_0}^t \dd \tau \  \sqrt{\dot{g}(\tau)^2 + \dot{h}(\tau)^2} 
\end{align}
where $\tau$ is a dummy integration variable and $\dot{h}(t) = \frac{d z}{dt}$, $\dot{g}(t) = \frac{d \rho}{dt}$.
Whenever the parameter $t$ can be eliminated to give $z = f(\rho)$ the equation above can be written as
\begin{align}\label{eq:arclenexpl}
s = \int_{\rho_0}^{\rho} \dd\mu \ \sqrt{1+\left( f'(\mu)\right)^2} 
\end{align}
where $f'(\rho) = \frac{df}{d\rho}$ and, as before, $\mu$ is a dummy integration variable.

To obtain the path of the waveguide we now need to find a curve $( \rho = g(t), \ z = h(t))$ (or $z = f(\rho)$)that satisfies  Eq.~\eqref{eq:F(s)} for a given $F(s)$ where we express $s$ as in Eq.~\eqref{eq:arclenparam} (or in Eq.~\eqref{eq:arclenexpl}) and where the angle $\theta$ satisfies 
\begin{align} \label{eq:customProfile}
\tan(\theta) = \frac{\dot{g}}{\dot{h}} \text{ or } \tan(\theta) =  \frac{1}{f'(\rho)}.
\end{align}
As a simple example consider the case where the target is to have $F(s) = \sin\left( \frac{2 \pi}{\Lambda} s \right)$ and we assume $d_0 = \varphi = 0$ and $k=1$. We claim that a path solution is
\begin{align}\label{eq:circle}
z = z_0+ r \sin t, \quad \rho = \rho_0 -r \cos t,
\end{align}
with $r = \Lambda/(2\pi)$. In this case we have that
\begin{align}
\tan \theta = \frac{\frac{d}{dt} (-a \cos t)}{\frac{d}{dt} (a \sin t)} = \tan t \text{ and } s = \Lambda t,
\end{align}
and we can now easily see that the curve in Eq.~\eqref{eq:circle} satisfies Eq.~\eqref{eq:F(s)}. This solution coincides with the original result by Yang et al.~\cite{yang2007enhanced,yang2007generating}.
While solving analytically the functional Eq.~\eqref{eq:F(s)} is often hard for more complex $F(s)$ the equation can be solved numerically as we discuss in Appendix~\ref{app:numerics}. Moreover, we provide numerical routines in Octave/Matlab and Python to numerically solve Eq.~\eqref{eq:F(s)} for arbitrary $F(s)$~\cite{salva_repo}. In the following examples, we use these numerical routines to obtain a Gaussian nonlinear profile for different crystal classes. Although, they can also be used to obtain other profiles aimed to other applications, e.g., nonlinear spectral holograms \cite{shiloh2012spectral, shapira2015nonlinear}, adiabatic processes in frequency conversion \cite{suchowski2014adiabatic,suchowski2008geometrical,karnieli2018fully}, phase matching in organic waveguides \cite{jazbinsek2019organic}.

\section{Simulations}\label{ref:simulations}
\subsection{Gaussian profile with modal phase matching}
\begin{figure} [h]
\centering
    \includegraphics[width=8.65cm]{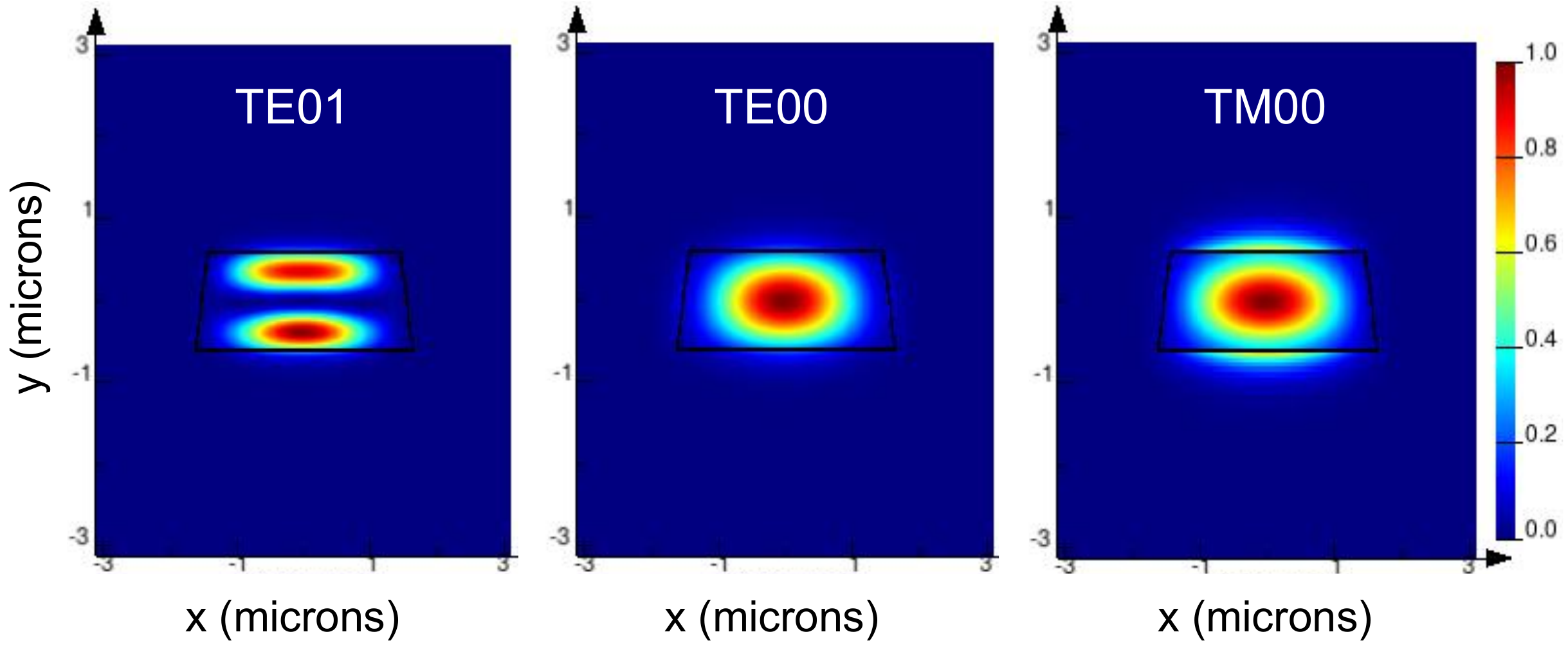}
    \caption{Electric field profiles for a CdS core and $\text{SiO}_2$ cladding, at 1550 nm for the TE01 mode and at 3100 nm for the TE00 and TM00 modes.}
    \label{fig:CdS100_modeProfile}
\end{figure}

In this section we design a waveguide with modal phase matching made of Cadmium Sulfide (CdS), with zincblende structure \cite{ichimura1999structural}, that has a Gaussian nonlinearity profile. The CdS wafer orientation can be in any  $\langle 100 \rangle$ direction.
Modal PM is the simplest case since no phase mismatch needs to be compensated. The waveguide geometry satisfies the phase matching condition, $k_p - k_s - k_i = 0$, where $k_l$ is the angular wavenumber of mode $l$, and the  group-velocity mismatch (GVM) condition 
$\left( v_{s}^{-1} - v_{p}^{-1} \right) / \left( v_{s}^{-1} - v_{i}^{-1} \right) > 0 $, where $v_l$ is the group velocity of mode $l$ \cite{graffitti2018design} and we label three modes participating in the three-wave mixing process signal (s), idler (i) and pump (p).
The channel width is 2.9~$\mu$m and the height is 1.2~$\mu$m. The core is surrounded by $\text{SiO}_2$.
The pump mode is TE01 at 1550 nm and the signal and idler modes are TM00 and TE00 at 3100 nm, respecting the energy conservation condition. 
The modal profiles are shown in Fig.~\eqref{fig:CdS100_modeProfile}. The simulations were conducted using the Finite Difference Eigenmode (FDE) solver from Lumerical Inc.

\begin{figure} 
\centering
    \includegraphics[width=8.6cm]{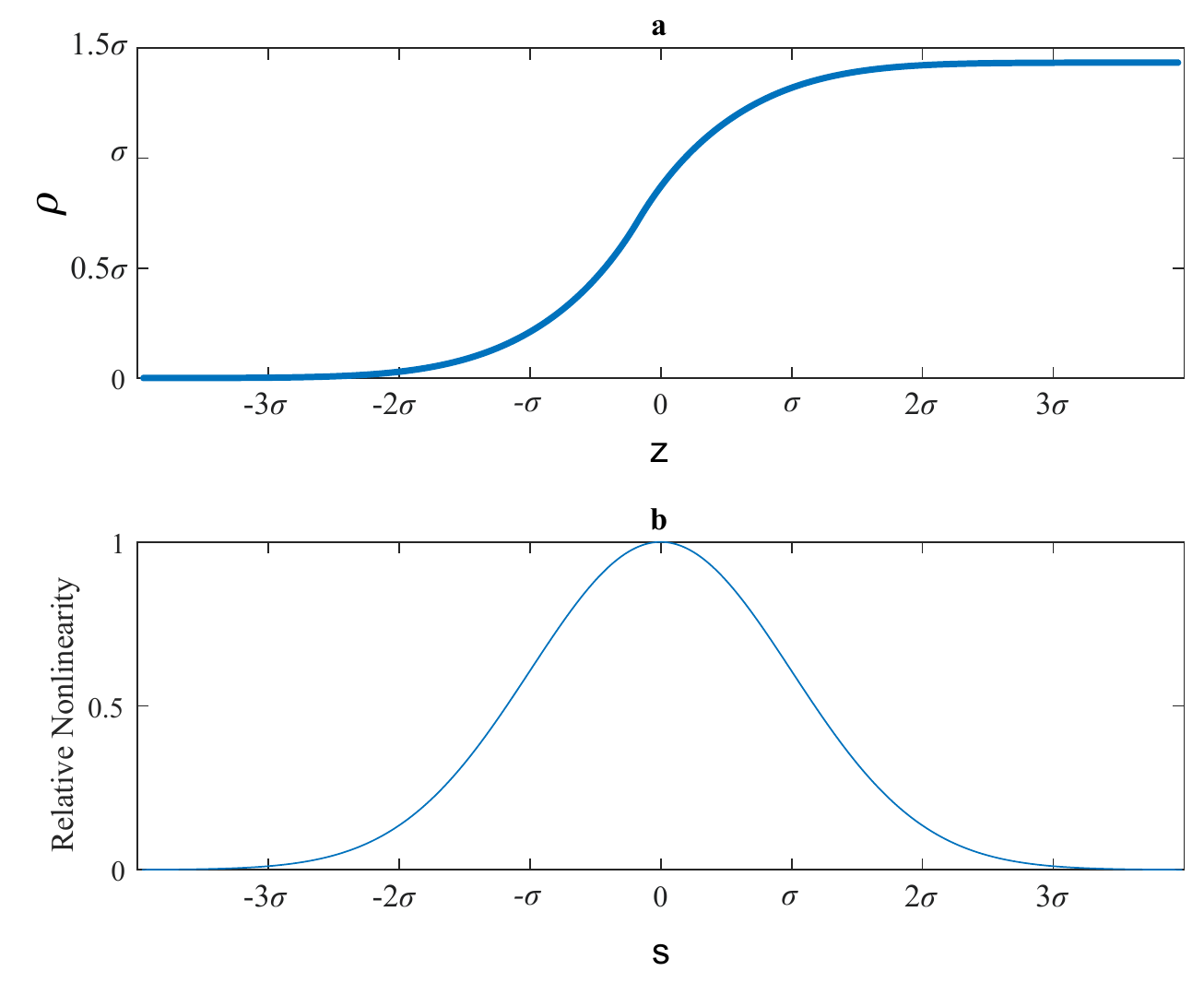}
    \caption{(a) Waveguide top-view  (b) Gaussian relative nonlinearity}
    \label{fig:gauss_waveguide}
\end{figure}

Zincblende belongs to the cubic $\bar{4}3m$ crystal class, in order to have a waveguide with effective nonlinear Gaussian profile, the waveguide orientation angle, at any point $s$, will be
\begin{align} \label{eq:theta43mGauss}
   \sin 2 \theta (s) = 
    \exp\left(-\tfrac12 \left[\frac{s}{L_\text{eff}} \right]^2 \right).
\end{align}
Using the methods described in Appendix~\ref{app:numerics} together with Eq.~\eqref{eq:theta43mGauss}, it is possible to solve numerically the waveguide path. For an arbitrary effective length $L_\text{eff}$, with initial conditions $z_0=-4L_\text{eff}$ and $\rho_0=0$, the waveguide path is shown in Fig.~\eqref{fig:gauss_waveguide}. 

The bandwidth of the signal and idler photons depends on the variance of the Gaussian profile which we labelled as the effective length $L_\text{eff}$. As shown in Fig.~\eqref{fig:gauss_waveguide}, the design of the waveguide is straightforward once the variance is fixed. When designing the waveguide, special attention must be paid to the bending radius, so that it does not alter the modal behaviour, in order to preserve phase matching. 
The minimum bending radius is linearly related to the standard deviation,
\begin{align} \label{eq:r_min_gauss}
    r_{\text{min}} = 2 L_\text{eff} ~.
\end{align}
Once calculated, the FDE analysis is recomputed to check that the phase matching condition is still accomplished. For example, for a waveguide of 2mm, $L_\text{eff}$ will be $250\mu$m and the minimum bending radius will be $500\mu$m. The losses caused by a $500\mu$m bending radius are negligible. The three modes overlap area is $3\text{mm}^2$. The average number of photons generated, calculated with Eq.~\eqref{eq:NdFinal}, is $4.7\times 10^{-15}~N_p$, where $N_p$ is the number of pump photons.

\subsection{Gaussian profile with angular phase matching (APM) - example 1}

The variation of the effective non-linearity respect to the orientation angle can be used to obtain a nonlinear sinusoidal profile and therefore add momentum to the phase mismatch in order to get momentum conservation. This technique named APM, was introduced to obtain quasi-PM in the context of microring resonators as shown in Ref.
\cite{yang2007enhanced, yang2007generating}. 
By carefully engineering the waveguide, a nonlinear profile made of a Gaussian modulated by a sine wave can be obtained (Eq.~\eqref{eq:gaussSin}),
\begin{align} 
\label{eq:gaussSin}
\frac{d_{\text{eff}}(s)}{d_0} = \exp\left(-\tfrac12 \left[\frac{s}{L_\text{eff}} \right]^2 \right) a \sin(2\pi s / \Lambda) ~.
\end{align}

\begin{figure} [h]
\centering
    \includegraphics[width=8.65cm]{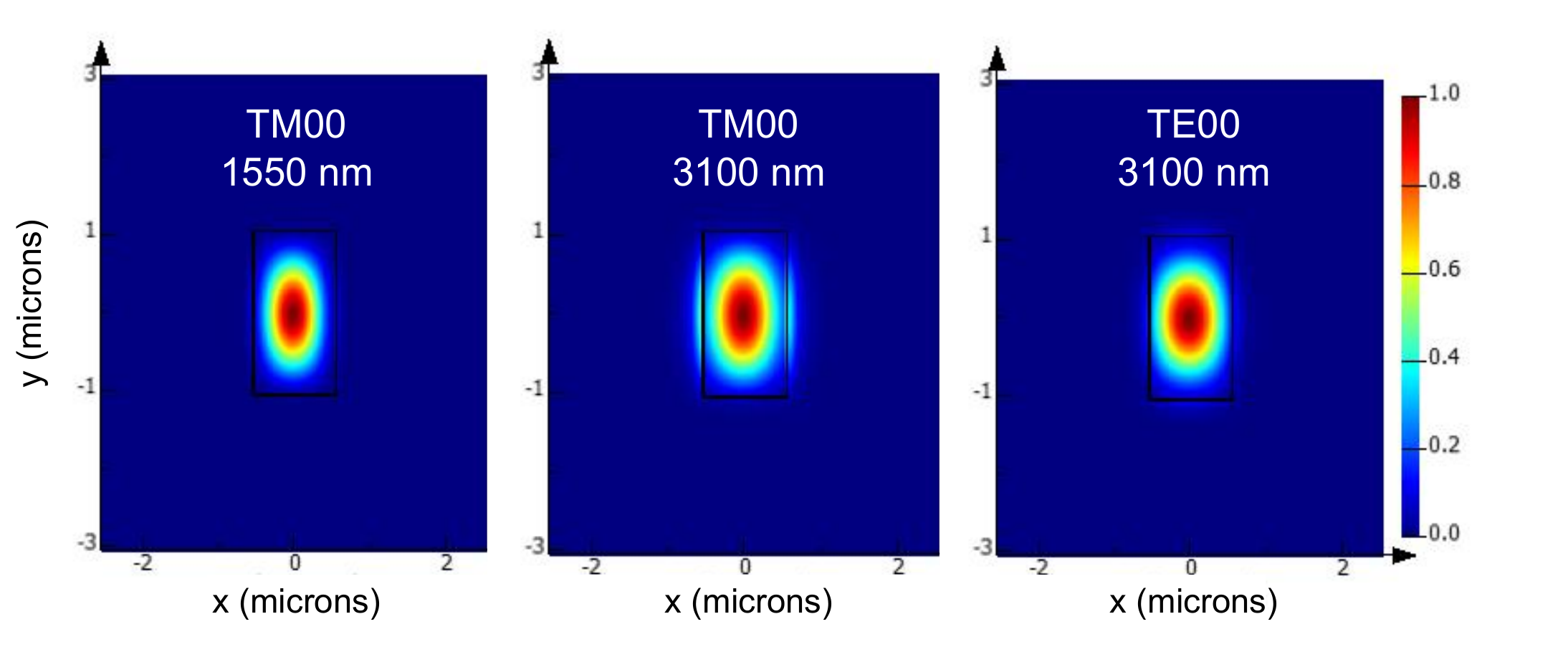}
    \caption{Electric field profiles for a GaP core and SiO2 cladding, at 1550~nm for the TE00 mode and at 3100~nm for the TE00 and TM00 modes.}
    \label{fig:modeProfiles_GaN}
\end{figure}

The value of $a$ in Eq. \eqref{eq:gaussSin} is a scalar with value between 0 and 1. The value of the minimum bending radius is increased by a factor of $1/a$ (Eq. \eqref{eq:r_min}),
\begin{align} \label{eq:r_min}
    r_{\text{min}} = \frac{\Lambda}{\pi a} = \frac{2}{\Delta k a}~,
\end{align}
and therefore the value of $a$ can be tuned so that the bending radius does not affect the phase mismatch by changing the effective refractive indices of the modes. 
For future waveguides designs, where bending losses is a limitation, $a$ can be tuned to decrease them.
In counterpart, decreasing $a$ to get a larger radius causes the effective non-linearity to decrease as well.
With this nonlinear profile, the PMF is then
\begin{align}
\begin{aligned}
    \int_{-\infty}^{\infty} \dd s \frac{\deff(s)}{d_0} e^{i \Delta k  s} = &i \sqrt{\frac{\pi }{2}} a L_\text{eff} e^{-\frac{L_\text{eff}^2}{2} (\Delta k - 2 \pi/\Lambda )^2}
      \\
    & - i \sqrt{\frac{\pi }{2}} a L_\text{eff} e^{-\frac{L_\text{eff}^2}{2} (\Delta k + 2 \pi/\Lambda )^2}  ~.
\end{aligned}
\end{align}
\begin{figure}
    \includegraphics[width=8.6cm]{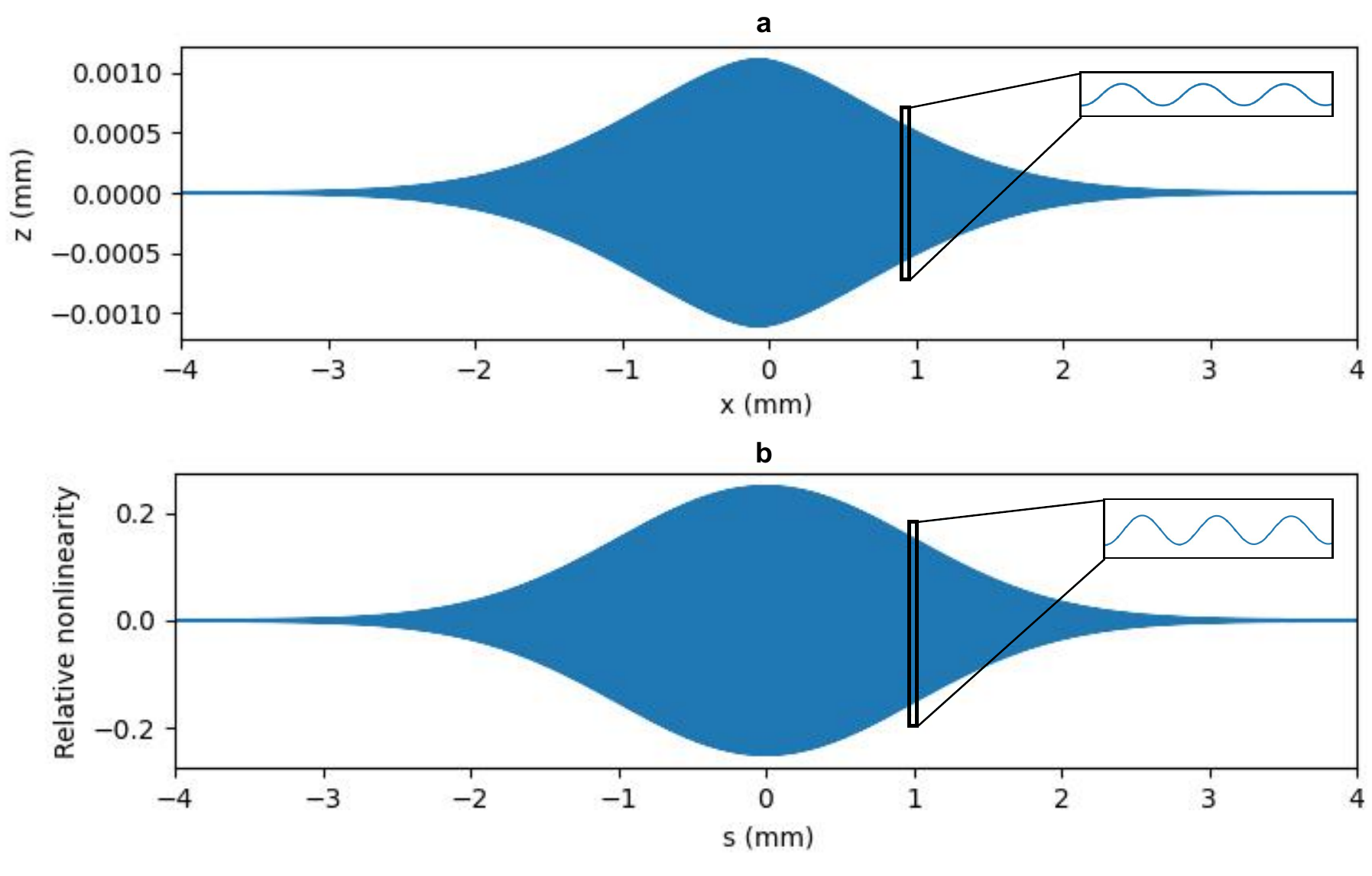}
    \caption{(a) Waveguide top-view  (b) Gaussian and sine relative nonlinearity}
    \label{fig:gaussSin_waveguide}
\end{figure}
An example of a waveguide with Gaussian profile and quasi-PM is shown in Fig.~\eqref{fig:gaussSin_waveguide}. The waveguide is made of Gallium Phosphide (GaP), which has zincblende crystal structure, the wafer can then be in any $\langle 100 \rangle$ crystal direction.
The channel width is 1.1~$\mu$m and the height is 2.1~$\mu$m.
The effective length $L_\text{eff}$ is fixed to $1~\text{mm}$.
The sine period is $\Lambda = 11.9~\mu$m, which leads to a minimum bending radius of $3.8~\mu$m. A waveguide with this bending radius does not guide the modes efficiently, hence, to get a waveguide with less than 0.1dB/cm loss, a bending radius of at least $14~\mu$m is needed. This bending radius can be achieved adjusting the parameter $a$ of Eq.~\eqref{eq:r_min} to 0.25 at the expense of decreasing the effective nonlinearity described in Eq.~\eqref{eq:gaussSin}.
The pump mode is TE00 at 1550~nm and the signal and idler modes are TM00 and TE00 at 3100~nm, to preserve the energy conservation condition. The mode profiles are shown in Fig.~\eqref{fig:modeProfiles_GaN}. The three modes correspond to the fundamental mode, which allows to maximize the mode overlap area to $1.79~\mu\text{m}^2$. The average number of photons generated, calculated with Eq.~\eqref{eq:NdFinal}, is $4\times 10^{-8}~N_p$. 
Notice that compared to the previous CdS example, the mode overlap area is  six orders of magnitude smaller, which implies an increase of six order of magnitude in the number of photons generated, hence the importance of having the three same order modes. 

\subsection{Gaussian profile with angular phase matching (APM) - example 2}
In this example, 
we get as target a nonlinearity profile that is the product of a Gaussian modulated by the absolute value of a sine wave, as opposed to the sine wave considered before,
\begin{align} \label{eq:gaussAbsSin}
\frac{d_{\text{eff}}(s)}{d_0} = \exp\left(-\tfrac12 \left[\frac{s}{L_\text{eff}} \right]^2 \right) a 
\left| \sin(2\pi s / \Lambda) \right| ~.
\end{align}
To calculate the PMF (Eq.~\eqref{eq:PMF_GAbsSin}), the Fourier transform of the absolute sine is needed (Eq.~\eqref{eq:FourierAbsSin}),
\begin{align} \label{eq:FourierAbsSin}
    \begin{aligned} 
        \left| \sin \left( \frac{2 \pi}{\Lambda} s\right) \right| 
        &= \frac{2}{\pi} - \frac{4}{\pi} \sum_{m=1}^{\infty} \frac{\cos\left( 2m \frac{2 \pi}{\Lambda} s\right)}{4m^2-1} \\
        &\approx \frac{2}{\pi} - \frac{4}{3\pi} \cos\left(\frac{4 \pi}{\Lambda} s\right) ~.\\
    \end{aligned}
\end{align}
Only the fastest term of the Fourier transform will be kept. Advantageously for the phase mismatch, the absolute sine frequency is twice of the pure sine. 
In counterpart, this comes at the expense of an effective nonlinearity 42\% smaller. The PMF is then
\begin{align} \label{eq:PMF_GAbsSin}
    \begin{aligned}
        \int_{-\infty}^{\infty} ds \frac{\deff(s)}{d_0} e^{i \Delta k s} = &
           \sqrt{\frac{\pi }{2}} \frac{4}{3\pi} a L_\text{eff} e^{-\frac{L_\text{eff}^2}{2} (\Delta k + 4 \pi/\Lambda )^2} \\
        & - \sqrt{\frac{\pi }{2}} \frac{4}{3\pi} a L_\text{eff} e^{-\frac{L_\text{eff}^2}{2} (\Delta k - 4 \pi/\Lambda )^2} ~.
    \end{aligned}
\end{align}
As for a sine wave, the value of the minimum bending radius is increased by a factor of $1/a$ (Eq.~\eqref{eq:r_min_Abs}), \begin{align} \label{eq:r_min_Abs}
    r_{\text{min}} = \frac{\Lambda}{\pi a} = \frac{4}{\Delta k a} ~,
\end{align}
and therefore the value of $a$ can be tuned so that the bending radius does not affect the phase mismatch.

\begin{figure} [h]
\centering
    \includegraphics[width=8.6cm]{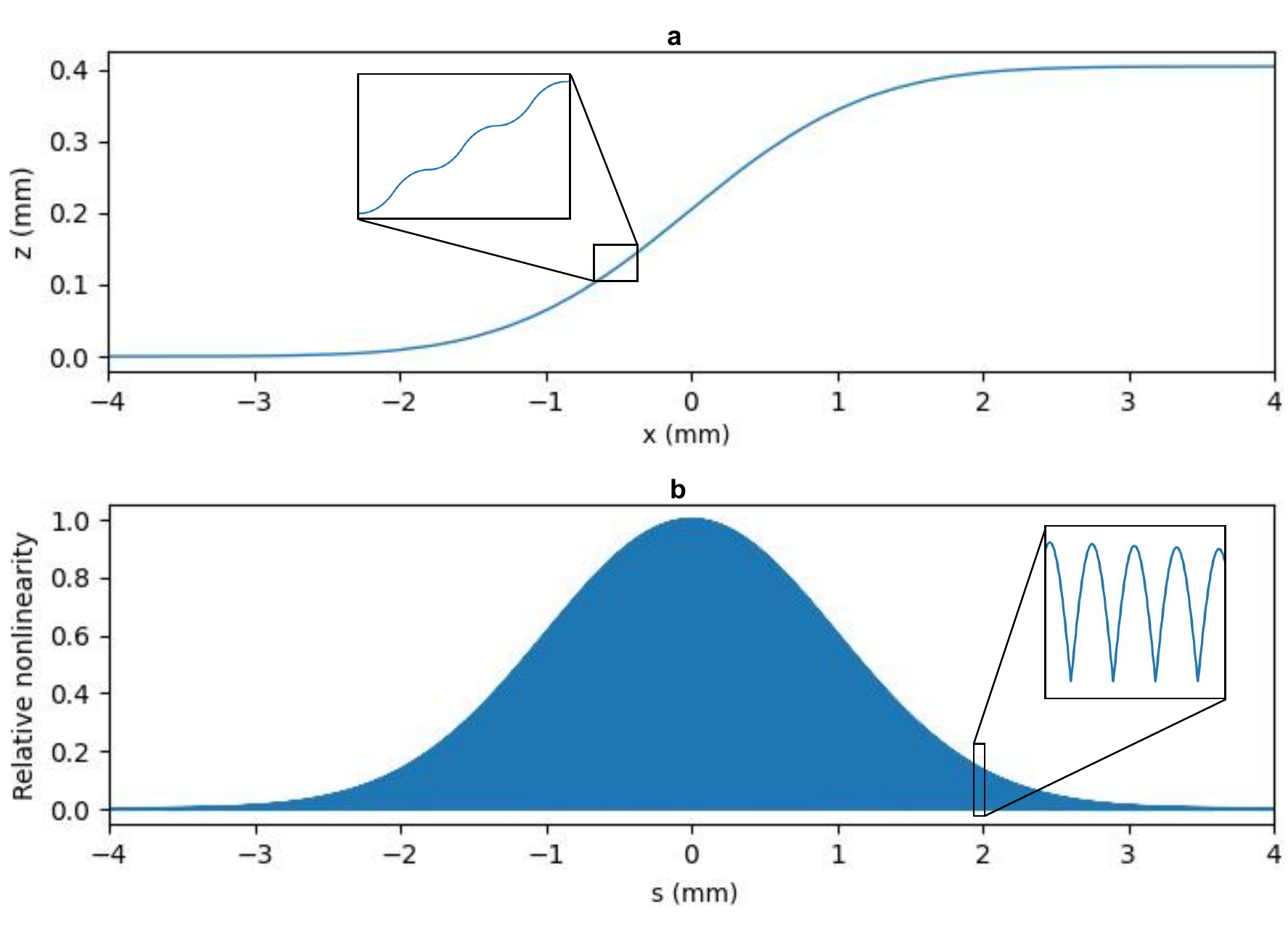}
    \caption{(a) Waveguide top-view  (b) Gaussian and absolute sine relative nonlinearity}
    \label{fig:gaussAbsSin_waveguide}
\end{figure}

In Fig.~\eqref{fig:gaussAbsSin_waveguide}, it is shown an example of waveguide geometry with Gaussian profile and quasi-PM with the same parameters of the previous example. Overall, the pure sine and the absolute sine are similar solutions to get quasi-PM. The pure sine has the advantage of having better effective non-linearity, however, as observed in the zoom of Fig.~\eqref{fig:gaussAbsSin_waveguide} there is no chance that the waveguide overlaps with itself contrary to Fig.~\eqref{fig:gaussSin_waveguide}. Therefore, for specific designs, it would be preferred to have an absolute sine nonlinear profile.

\section{Conclusion and Discussion}\label{ref:conclusions}

We have demonstrated that high-purity photon pairs can be generated in a single crystal by numerically designing the waveguide 2D path. The technique we propose takes advantage of the tensor nature of the nonlinear susceptibility. Since this quantity has not only magnitude but also direction, we can alter the susceptibility seen by propagating fields by locally altering the propagation \emph{direction}.
Our technique increases the material platforms suitable for the generation of high-purity heralded single-photon generation, since neither OP and PP is required, it is not susceptible to random duty cycle errors and variations in the domain width. It is then an alternative for materials for which it is hard to do OP or PP.
The fabrication process is comparatively straightforward, however, for this method to work properly, it is necessary that the waveguide is composed of a single crystal, free from twin boundaries and dislocations which can alter the crystalline orientation and compromise the performance of the nonlinear process.
The results shown in Table~\ref{tab:summaryAPM} and Eq.~\eqref{eq:customProfile} will allow other researchers to create custom effective nonlinearity profiles beyond Gaussian functions.
In addition, quasi-PM is compatible with having a Gaussian nonlinear profile enlarging the range of possible materials where momentum conservation and GVM condition are satisfied. However, for large values of phase mismatch, the bending radius is quite small. Having small bending radius makes the modal behaviour to vary and it increase the losses which makes, for some cases, the design not feasible in practice. Also, for birefringent crystals, the refractive index changes with the waveguide orientation and then the phase mismatch and GVM change with it. The width of the waveguide can be tuned in order to preserve the phase mismatch, however, the GVM will slightly vary, and therefore it will need further designing to get perfect purity.

\begin{acknowledgments}
The authors would like to thank Hossein Seifoory, Marc-Antoine Bianki, Régis Guertin, Cédric Lemieux-Leduc, Ahmed Bahgat and Thomas Lacasse
for helpful suggestions and discussions. N. Q. and Y.-A. P. acknowledge support from the Ministère de l’Économie et de l’Innovation du Québec and the Natural Sciences and Engineering Research Council of Canada.
\end{acknowledgments}

\appendix

\section{Numerical solution of the path equation}\label{app:numerics}
In this appendix we introduce a simple algorithm to obtain the path $(\rho,z)$ given a target nonlinearity profile $F(s)$ along the propagation direction $s$. We discretize the path in step of size $\Delta z$.
Given an initial point $(\rho_0,z_0)$ the algorithm generate the next point in the path by using the knowledge of the desired nonlinearity value as follows:
\begin{algorithmic}
\State $\rho[0] \gets \rho_0$
\State $z[0] \gets z_0$
\State $s \gets 0$
\While{$n \leq N$}
    \State $z[n] \gets z[n-1] + \Delta z$
    \State $s[n] \gets s[n-1] + \sqrt{(\Delta z)^2+(\rho[n-1]-\rho[n-2])^2}$
    \State $\theta \gets \left( \sin^{-1}( F(s) - d_0/d)-\varphi \right)/k$
    \State $\rho[n] \gets \rho[n-1] +\tan (\theta) \times \Delta z$
    \State $n \gets n+1$
\EndWhile
\end{algorithmic}

\section{Average number of photons generated}
Here we follow the approach from Ref.~\cite{yang2008spontaneous} to calculate the average number of photons generated by SPDC. We will develop here the specific case in which two spectrally uncorrelated photons are generated. The general equation of the average number of photons generated is
\begin{align} \label{eq:N_D1}
    &N_D = \frac{\hbar \left|\chi^{(2)}\right|^2}{4 \pi \Bar{n}^6 \varepsilon_0 A} N_P  \left[ \frac{\omega_{i0}}{v_i} \frac{\omega_{s0}}{v_s} \frac{\omega_{p0}}{v_p} \right] \times \\
    & \times \int^\infty_0 \dd \omega_i  
    \int^\infty_0  \dd \omega_s
    \left|
    \phi_p(\omega_s+\omega_i)
    \int_{-\infty}^{\infty} \dd s \frac{\deff(s)}{d_0} e^{i \Delta k \ s} 
    \right|^2, \nonumber
\end{align}
where $N_P$ is the average number of pump photons, $A$ is the effective area and $\phi_p$ is $L^2$ normalized pump pulse amplitude.
In the last equation we have evaluated all the slowly varying quantities depending on the frequencies at the central frequencies of the respective mode, thus for example replacing $\frac{d k(\omega_j)}{d \omega_j} \to \frac{d k(\omega_j)}{d \omega}|_{\omega_j = {\omega}_{j0}} \equiv  \frac{1}{v_j}$ for $j \in \{s,i,p\}$.

We will consider a Gaussian pump pulse 
\begin{align}
    \phi_p(\omega_p) = \left( \sqrt{\pi} \sigma_p \right)^{-\tfrac{1}{2}}
    \exp\left[ -\frac{1}{2} \left( \frac{\omega_p - \omega_{p0}}{\sigma_p} \right) ^2 \right]  ~,
\end{align}
and a waveguide with dispersion modelled by a second order Taylor expansion around the centre frequency
\begin{align}
    k &= 
    k_{i0} + \frac{\omega_i - \omega_{i0} }{v_{i}} + \frac{1}{2}\frac{\partial^2 k}{\partial \omega_i^2}(\omega_i - \omega_{i0})^2
\end{align}
where $\omega_{i0}$ is the central frequency of mode $i$ and $v_i$ the group velocity of mode $i$.
We have checked that for the waveguides proposed here the second order coefficient has negligible effect on the integrals, so it can be removed in the further development.
The effective nonlinear susceptibility, $\deff(s)/d_0$, has a Gaussian profile
modulated by a sine wave to get phase matching (Eq.~\eqref{eq:gaussSin}). In Eq.~\eqref{eq:N_D1} we have then the double integral of two Gaussians. To make them spectrally indistinguishable we need to force the Gaussian cross-term to vanish
\begin{align}
\label{eq:crossterm}
    \frac{1}{\sigma_p^2} = -
    L_\text{eff}^2 \left(\frac{1}{v_p}-\frac{1}{v_s} \right)\left(\frac{1}{v_p}-\frac{1}{v_i}\right) ~.
\end{align}
The integral is then straightforward. The average number of photons generated by a waveguides having a Gaussian nonlinear profile is
\begin{align} 
\label{eq:NdFinal}
    N_D = 
    \frac{\sqrt{\pi}\hbar \left\{\chi^{(2)}\right\}^2  \bar{\omega}_s \bar{\omega}_i \bar{\omega}_p L_\text{eff}}{8 \varepsilon_0 \Bar{n}^6 v_p A} \frac{N_P}{|v_s - v_i|}  ~,
\end{align}
consistent with the results in Appendix B of Ref.~\cite{triginer2020understanding}.

\nocite{*}

\bibliography{apssamp}

\end{document}